\documentclass[aip,reprint]{revtex4-1}

\usepackage{graphicx}
\usepackage{dcolumn}
\usepackage{bm}
\usepackage{amssymb}
\usepackage{amsmath}
\usepackage[T1]{fontenc}
\usepackage{textcomp}
\newcommand{\units}[3][]{$#1\mathrm{#2\,#3}$}
\newcommand{\mum}{\mbox{\textmu m}}
\newcommand{\Happl}{H_\mathrm{appl}}
\newcommand{\Hind}{H_\mathrm{ind}}
\newcommand{\Htot}{H_\mathrm{total}}
\newcommand{\Hfmr}{H_\mathrm{FMR}}
\newcommand{\Ms}{M_\mathrm{s}}
\newcommand{\hp}{h_\mathrm{p}}
\newcommand{\hth}{h_\mathrm{th}}
\newcommand{\hmin}{h_\mathrm{th,min}}
\newcommand{\jc}{j_\mathrm{c}}
\newcommand{\wrx}{\Gamma}

\begin{document}

\title{Spin-transfer torque based damping control of parametrically excited spin waves in a magnetic insulator}

\author{V. Lauer}
\affiliation{Fachbereich Physik and Landesforschungszentrum OPTIMAS, Technische Universit\"at
Kaiserslautern, 67663 Kaiserslautern, Germany}
\author{D.~A. Bozhko}
\affiliation{Fachbereich Physik and Landesforschungszentrum OPTIMAS, Technische Universit\"at
Kaiserslautern, 67663 Kaiserslautern, Germany}
\affiliation{Graduate School Materials Science in Mainz, Gottlieb-Daimler-Strasse 47, 67663 Kaiserslautern, Germany}
\author{T. Br\"acher}
\affiliation{Fachbereich Physik and Landesforschungszentrum OPTIMAS, Technische Universit\"at
Kaiserslautern, 67663 Kaiserslautern, Germany}
\author{P. Pirro}
\altaffiliation[Present address: ]{Institut Jean Lamour, Universit\'{e} Lorraine, CNRS, 54506 Vandoeuvre-l\`{e}s-Nancy, France}
\affiliation{Fachbereich Physik and Landesforschungszentrum OPTIMAS, Technische Universit\"at
Kaiserslautern, 67663 Kaiserslautern, Germany}
\author{V.~I. Vasyuchka}
\affiliation{Fachbereich Physik and Landesforschungszentrum OPTIMAS, Technische Universit\"at
Kaiserslautern, 67663 Kaiserslautern, Germany}
\author{A.~A. Serga}
\affiliation{Fachbereich Physik and Landesforschungszentrum OPTIMAS, Technische Universit\"at
Kaiserslautern, 67663 Kaiserslautern, Germany}
\author{M.~B. Jungfleisch}
\altaffiliation[Present address: ]{Materials Science Devision, Argonne National Laboratory, Argonne, Illinois 60439, USA}
\affiliation{Fachbereich Physik and Landesforschungszentrum OPTIMAS, Technische Universit\"at
Kaiserslautern, 67663 Kaiserslautern, Germany}
\author{M. Agrawal}
\affiliation{Fachbereich Physik and Landesforschungszentrum OPTIMAS, Technische Universit\"at
Kaiserslautern, 67663 Kaiserslautern, Germany}
\author{Yu.~V. Kobljanskyj}
\affiliation{Faculty of Radiophysics, Electronics and Computer Systems, Taras Shevchenko National University of Kyiv, 01601 Kyiv, Ukraine}
\author{G. A. Melkov}
\affiliation{Faculty of Radiophysics, Electronics and Computer Systems, Taras Shevchenko National University of Kyiv, 01601 Kyiv, Ukraine}
\author{C. Dubs}
\affiliation{Innovent e.V., Pr\"ussingstra\ss e 27B, 07745 Jena, Germany}
\author{B. Hillebrands}
\affiliation{Fachbereich Physik and Landesforschungszentrum OPTIMAS, Technische Universit\"at
Kaiserslautern, 67663 Kaiserslautern, Germany}
\author{A.~V. Chumak}
\affiliation{Fachbereich Physik and Landesforschungszentrum OPTIMAS, Technische Universit\"at
Kaiserslautern, 67663 Kaiserslautern, Germany}

\date{\today}

\begin{abstract}
The damping of spin waves parametrically excited in the magnetic insulator Yttrium Iron Garnet (YIG) is controlled by a dc current passed through an adjacent normal-metal film. The experiment is performed on a macroscopically sized YIG(\units{100}{nm})/Pt(\units{10}{nm}) bilayer of \units{4 \times 2}{mm^2} lateral dimensions. The spin-wave relaxation frequency is determined via the threshold of the parametric instability measured by Brillouin light scattering (BLS) spectroscopy. The application of a dc current to the Pt film leads to the formation of a spin-polarized electron current normal to the film plane due to the spin Hall effect (SHE). This spin current exerts a spin transfer torque (STT) in the YIG film and, thus, changes the spin-wave damping. Depending on the polarity of the applied dc current with respect to the magnetization direction, the damping can be increased or decreased. The magnitude of its variation is proportional to the applied current. A variation in the relaxation frequency of $\pm7.5 \% $ is achieved for an applied dc current density of \units{5 \cdot 10^{10}}{A/m^2}. 

\end{abstract}


\maketitle

The injection of a spin current into a magnetic film can generate a spin-transfer torque (STT) that acts on the magnetization collinearly to the damping torque.\cite{Slonczewski1996,Berger1996, Slavin-Review} Thus, it can be used for tuning of the damping of a magnetic film\cite{KrivorotovScience307,AndoPRL101, DemidovAPL99,HamadehPRL113} as well as for the excitation of magnetization precession in the film.\cite{TsoiPRL80, DemidovNatMater9, MadamiNatNanotech6, DemidovNatMater11,ColletArxiv}
In the first experimental realizations, a dc charge current was sent through an additional magnetic layer with a fixed magnetization direction in order to generate a spin-polarized current.\cite{TsoiPRL80,KrivorotovScience307,DemidovNatMater9, MadamiNatNanotech6} A different way to generate a spin current is based on the spin Hall effect (SHE)\cite{DyakonovPLA35,HirschPRL83} caused by spin-dependent scattering of electrons in a non-magnetic metal with large spin-orbit interaction.\cite{AndoPRL101, DemidovAPL99, DemidovNatMater11} One of the advantages of the SHE is that it does not require a dc current in the magnetic layer and, thus, allows for the application of a STT to a magnetic dielectric such as yttrium iron garnet (YIG),\cite{KajiwaraNature464} which is of particular interest for magnon spintronics\cite{KruglyakJPhysD43,SergaJPhysD43,ChumakNatPhys11} due to its extremely small damping parameter.\cite{AllivyKellyAPL103,PirroAPL104,OnbasliAPLMater2, ChangMagnLettIEEE5,JungfleischJAP117} Moreover, a great advantage of the SHE is that a STT can be applied not only locally but to a large area of a magnetic film\cite{AndoPRL101} and, thus, can be potentially used to compensate the damping in a whole complex magnonic circuit.\cite{ChumakNatPhys11}

Up to now, the auto-oscillatory regime and the magnetization precession generation has only been reached in patterned structures.\cite{DemidovNatMater11,ColletArxiv,DuanNatCommun5} Nonlinear multi-magnon scattering phenomena are assumed to be the reason that disturbs the generation process in un-patterned stuctures. At the same time, nonlinear scattering should not affect the SHE-STT-based damping compensation since spin-wave amplitudes in this case are much smaller. However, most experimental studies concerning this phenomena using metallic samples\cite{KrivorotovScience307, DemidovAPL99} as well as YIG structures\cite{HamadehPRL113,ColletArxiv} were performed with laterally-confined nano- or micro-structures.

Here, we use a YIG/Pt bilayer of macroscopic size to investigate SHE-STT damping variation. The measured variation of the damping is proportional to the applied dc current and no influence of multi-scattering magnon processes is observed.

The investigated sample consists of a YIG/Pt bilayer with macroscopic lateral dimensions of \units{4}{mm} $\times$ \units{2}{mm}, and film thicknesses of \units[t_\mathrm{YIG}=]{100}{nm} and \units[t_\mathrm{Pt}=]{10}{nm}. The YIG film is grown by liquid phase epitaxy on a gadolinium gallium garnet (GGG) substrate and the Pt film is deposited afterwards using plasma cleaning and RF sputtering, as described in Ref. \onlinecite{PirroAPL104}.
Measurements of the ferromagnetic resonance (FMR) linewidth and of the inverse SHE induced by spin pumping in a wide frequency range yield the Gilbert damping parameters \units[\alpha_\mathrm{YIG}=]{(1.3\pm 0.1)\cdot 10^{-4}}{} for the bare YIG film and \units[\alpha_\mathrm{YIG/Pt}=]{(5.3\pm 0.1)\cdot 10^{-4}}{} for the YIG/Pt bilayer. In addition, a saturation magnetization \units[\mu_0 \Ms=]{(173 \pm 1)}{mT}, an inhomogeneous broadening \units[\mu_0 \Delta H_0 =]{0.26}{mT}, a resistivity of the Pt film  \units[\rho_\mathrm{Pt}=]{1.475 \cdot 10^{-7}}{\Omega m}, and an effective spin mixing conductance \units[g_\mathrm{eff}^{\uparrow\downarrow}=]{3.68 \cdot 10^{18}}{m^{-2}} are determined at room temperature.

\begin{figure}[]
\begin{center}
\scalebox{1}{\includegraphics[width=0.4 \textwidth, clip]{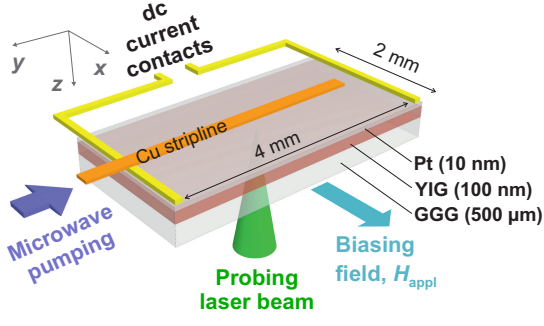}}
\end{center}
\caption{\label{Fig_sample}(color online) Sketch of the sample and the setup in the parallel pumping geometry used for the parametric excitation of spin waves and their detection by BLS spectroscopy. The externally applied biasing field and the pumping field in the examined area of the sample are both along the $x$-direction, the charge current is applied in-plane along the $y$-direction.}
\end{figure}

The investigation of the SHE-STT damping variation is based on the analysis of the threshold of the parametric instability. Figure~\ref{Fig_sample} shows a sketch of the experimental arrangement, in which the parametric instability threshold is measured in the parallel pumping geometry.\cite{SergaJPhysD43,SandwegPRL106,EdwardsPRB86} An external biasing field $\Happl$ magnetizes the YIG film in-plane along the $x$-axis. The parametric excitation of spin waves is achieved by an alternating pumping magnetic field $\hp$ oriented parallel to $\Happl$. For this purpose, a microwave signal at a fixed frequency of \units[f_\mathrm{p}=]{14}{GHz} is applied to a \units{50}{\mum} wide Cu microstrip antenna placed on top of the sample. A \units{10}{\mum} thick polyethylene interlayer separates the antenna from the sample electrically.
In our experiment, spin waves at half of the pumping frequency (\units[f_\mathrm{sw}=f_\mathrm{p}/2=]{7}{GHz}) are excited as soon as the pumping field amplitude $\hp$ overcomes a critical threshold value $\hth$. The detection of these spin waves is realized by means of BLS spectroscopy.\cite{BookBLS} The incident probing laser beam in our experiment, which accesses the YIG film through the GGG substrate (see Fig.~\ref{Fig_sample}), is always perpendicular to the film plane and only spin waves with wavenumbers in a range of \units[k \lesssim]{10^4}{rad/cm} are detected.\cite{SandwegRSI81} Since the threshold $\hth$ depends on the biasing field, both parameters, $\Happl$ and $\hp$, are varied in each measurement.
Furthermore, Au wires are mounted to the edges of the sample by silver conductive adhesive to apply an in-plane dc current to the Pt film along the $y$-axis. Since the dc current is perpendicular to $\Happl$, the SHE generates an out-of-plane spin current (along the $z$-direction) in the Pt film that exerts a STT on the YIG magnetization at the YIG/Pt interface. A maximal current of \units{1}{A} is used, corresponding to a current density of \units[\jc=]{5 \cdot 10^{10}}{A/m^2}. 
In order to reduce the influence of Joule heating in the Pt film, the experiment is performed in the pulsed regime with a pulse duration of \units{10}{\mbox{\textmu s}} and a repetition time of \units{1}{ms}. All measurements presented here are performed at room temperature.

\begin{figure}[h]
\begin{center}
\scalebox{1}{\includegraphics[width=0.42 \textwidth, clip]{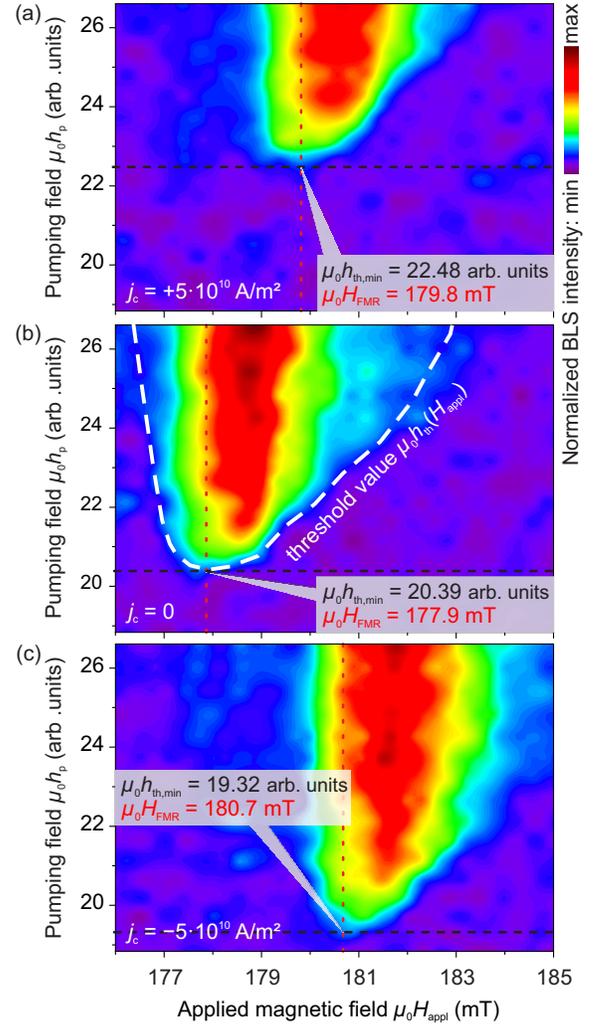}}
\end{center}
\caption{\label{Fig_BLS_butterfly}(color online) BLS intensity of parametrically pumped spin waves at \units{7}{GHz} while scanning the biasing field $\Happl$ and the pumping field $\hp$, for different applied current densities (a) \units[\jc =]{+5\cdot 10^{10}}{A/m^2}, (b) \units[\jc =]{0}{}, (c) \units[\jc =]{-5\cdot 10^{10}}{A/m^2}. The dashed white line in (b) represents the butterfly curve for spin waves with \units[k \lesssim]{10^4}{rad/cm}.
}
\end{figure}

Figures~\ref{Fig_BLS_butterfly}(a)-(c) exemplary show BLS intensities of spin waves at \units[f_\mathrm{sw}=]{7}{GHz} when the pumping field $\hp$ and the biasing field $\Happl$ (in $+x$-direction) are swept, measured for three different current densities $\jc$ (in $+y$-direction). The applied current densities are (a) \units[\jc =]{+5\cdot 10^{10}}{A/m^2}, (b) reference measurement \units[\jc =]{0}{}, (c) \units[\jc =]{-5\cdot 10^{10}}{A/m^2}. Blue colored areas in the intensity graphs correspond to the case $\hp < \hth$ when no spin waves are parametrically excited. 

The density $n(\vec{k},t)$ of parametrically excited magnons of a certain $\vec{k}$-vektor at the time $t$ in the vicinity of the threshold is given by\cite{Melkov_Book,Lvov_Book}
\begin{equation}
n(\vec{k},t) = n_0(\vec{k}) \cdot \mathrm{exp} \left[ {-2 \left( \wrx(\vec{k})-|V(\vec{k})| \mu_0 \hp \right) t} \right],
\label{eq:pumping}
\end{equation}
where $n_0$ is the initial magnon density (thermal level), $\wrx$ is the relaxation frequency, $V(\vec{k})$ is the coupling parameter between magnons and the pumping field, and $\mu_0$ is the vacuum permeability.
The magnon density grows exponentially, as soon as the pumping field overcomes a critical threshold value of $\hp\geq\hth= \mathrm{min} \left\lbrace \dfrac{\wrx(\vec{k})}{\mu_0 |V(\vec{k})|} \right\rbrace$ for one mode with the wavevector $\vec{k}$.
For a fixed pumping frequency, the $\hth$ value strongly depends on the biasing field $\Happl$, which determines the wavevector of the available spin waves. The diagram of $\hth$ vs. $\Happl$, called butterfly curve,\cite{Melkov_Book} exhibits a minimum $\hmin$ at the resonant field $\Hfmr$ which corresponds to the excitation of spin waves with $k \rightarrow 0$. The white dashed line in Fig.~\ref{Fig_BLS_butterfly}(b) represents only the part of the butterfly curve (for the BLS accessible wavenumbers) in a small range of the biasing field around its minimum. 
For the further study, only the variations of $\hmin$ and $\Hfmr$, related to $k \rightarrow 0$ spin waves, are analyzed as functions of the applied current density $\jc$.

\begin{figure}[]
\begin{center}
\scalebox{1}{\includegraphics[width=0.49 \textwidth, clip]{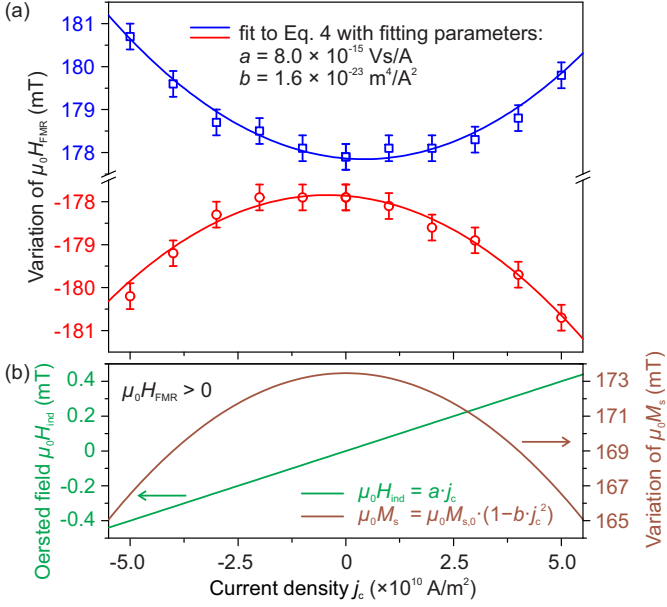}}
\end{center}
\caption{\label{Fig_shift_Hext} (color online) (a) Obtained $\Hfmr$ values (in $+x$-direction) at $\hmin$ as function of $\jc$. The solid lines are fits to Eq.~\ref{eq:solution}. (b) The induced Oersted field (in $+x$-direction) and the variation of the saturation magnetization as functions of $\jc$, when $\Hfmr > 0$ (in $+x$-direction).}
\end{figure}

First, the variation of the resonant biasing field $\Hfmr$ corresponding to the minimal threshold pumping field $\hmin$ is investigated. Figure~\ref{Fig_shift_Hext}(a) presents all $\Hfmr$ values obtained from BLS measurements (as shown, e.g., in Figs.~\ref{Fig_BLS_butterfly}(a)-(c)) for both directions of the biasing field and  both directions of the current. $\Hfmr$ evidently shifts always to values of higher magnitude with increasing $\jc$, regardless of the current direction, but the magnitudes of the shift are different for opposite current directions.
The observed behavior can be understood by taking into account the contributions of two effects:
(i) The dc current in the Pt induces an Oersted field $\Hind$ collinear to $\Happl$, which is proportional to $\jc$. Thus, the total magnetic field in the YIG film reads
\begin{equation}
\Htot = \Happl + \Hind = \Happl + \dfrac{a}{\mu_0} \cdot \jc , \\
\label{eq:def-a}
\end{equation}
where the proportionality constant $a$ is introduced as a fitting parameter accounting for the sample geometry. 
(ii) Joule heating in the Pt film leads to a temperature increase in the bilayer, and decreases the saturation magnetization of the YIG film following
\begin{equation}
\Ms(\jc) = \Ms(0) \cdot (1- \beta \cdot \Delta T) = \Ms(0) \cdot (1-b \cdot {\jc}^2).
\label{eq:def-b}
\end{equation}
Here, \units[\mu_0 \Ms(0)=]{(173 \pm 1)}{mT} is the initial saturation magnetization at room temperature (\units{300}{K}) at \units[\jc =]{0}{}. The factor \units[\beta=]{(2.2 \pm 0.1)\cdot 10^{-3}}{K^{-1}} accounts for the change of $\Ms$ with a temperature change of $\Delta T$,\cite{AndersonPRA134} which in turn is assumed to depend on the applied current as $\Delta T = b / \beta \cdot {\jc}^2$ by introducing the fitting parameter $b$. Using Eq.~\ref{eq:def-a}, Eq.~\ref{eq:def-b} and the Kittel equation ($k \rightarrow 0$)
$f_\mathrm{sw} = \gamma \mu_0 \sqrt{(\Hfmr + \Hind) (\Hfmr + \Hind + \Ms)}$, the $\Hfmr$ values in Fig.~\ref{Fig_shift_Hext}(a) can be fitted by
\begin{equation}
\begin{split}
\pm \mu_0 |\Hfmr|  = - a \cdot \jc \mp \dfrac{1}{2} \mu_0 \Ms(0) \cdot (1-b \cdot {\jc}^2) \\
\pm \sqrt{ \left( \dfrac{f_\mathrm{sw}}{\gamma} \right) ^2 + \left( \dfrac{1}{2} \mu_0 \Ms(0) \cdot (1-b \cdot {\jc}^2) \right) ^2},
\end{split}
\label{eq:solution}
\end{equation}
where \units[\gamma=]{28}{GHz/T} denotes the gyromagnetic ratio. The sign of $\pm\mu_0|\Hfmr|$ indicates the field polarity ("$+$" corresponds to the $+x$-direction).
The solid lines in Fig.~\ref{Fig_shift_Hext}(a) represent the fit according to Eq.~\ref{eq:solution} yielding the fitting parameters \units[a=]{8.0 \cdot 10^{-15}}{Vs/A} and \units[b=]{1.6 \cdot 10^{-23}}{m^4/A^2}. Based on the values of $a$ and $b$ the variation of the Oersted field $\mu_0 \Hind$ and the magnetization $\mu_0 \Ms$ with $\jc$ are shown in Fig.~\ref{Fig_shift_Hext}(b) for the case of $\mu_0 \Hfmr >0$. The highest applied current density of \units[\jc =]{\pm 5 \cdot 10^{10}}{A/m} decreases $\mu_0 \Ms$ by \units{\approx 7.0}{mT}, yielding a corresponding temperature increase of \units[\Delta T \approx]{18}{K} for the YIG film.
The change of $\mu_0 \Hind$ for the maximal $\jc$ is \units{\approx \pm 0.4}{mT}. This magnitude agrees with the theoretically expected Oersted field given by
\units[(\mu_0 /2) \jc t_\mathrm{Pt}\approx]{\pm 0.3}{mT}, if the sample width is considered to be much larger than the film thickness $t_\mathrm{YIG}$. This suggests that no STT-based variation of the saturation magnetization of YIG is observed in our experiment.\cite{DemidovAPL99}

\begin{figure}[]
\begin{center}
\scalebox{1}{\includegraphics[width=0.45 \textwidth, clip]{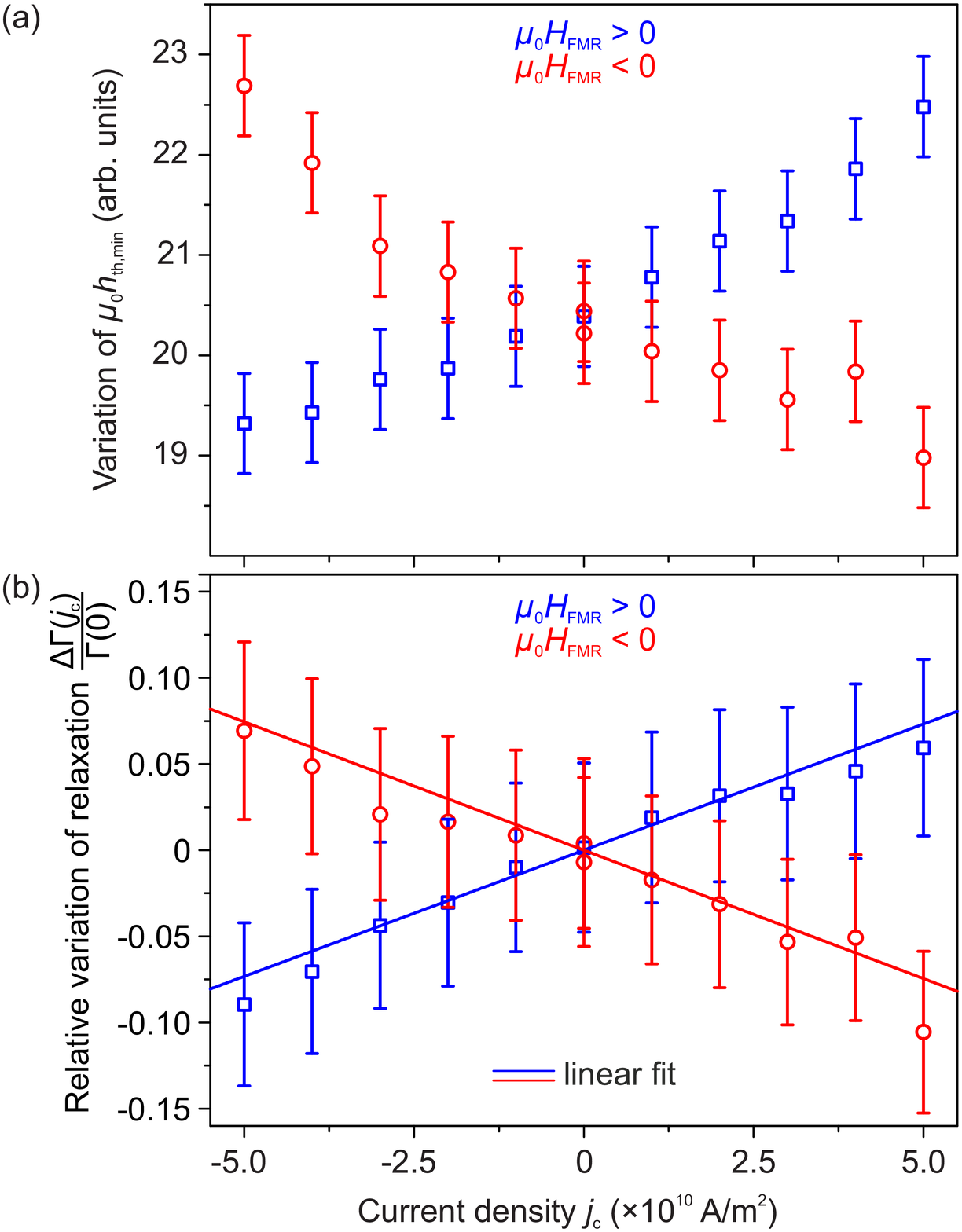}}
\end{center}
\caption{\label{Fig_shift_omega}(color online) (a) Measured minimum threshold values $\hmin$ extracted from BLS as a function of $\jc$ for positive and negative $\Hfmr$. (b) Relative change of the relaxation frequency according to Eq.~\ref{eq:wR-change}. The solid lines are linear fits.}
\end{figure}

Figure~\ref{Fig_shift_omega}(a) presents the minimal threshold pumping field $\hmin$ as function of $\jc$, extracted from data such as shown in Figs.~\ref{Fig_BLS_butterfly}(a)-(c). The shift of $\hmin$ obviously depends on the orientation of $\jc$ with respect to $\Happl$ and can be understood in terms of the SHE-STT effect. According to our experimental geometry, the direction of the SHE-generated spin current in the Pt film is given by the unity vector $(\vec{\jc} \times \vec{H}_\mathrm{appl})/|\vec{\jc} \times \vec{H}_\mathrm{appl}|$.\cite{SchreierJPD48} If the spin current, for example, is oriented in $+z$-direction (from Pt to YIG), the damping in the YIG film and consequently $\hmin$ decreases (compare Fig.~\ref{Fig_BLS_butterfly}(b) with (c)). In the case of a reversed spin current direction (from YIG to Pt), $\hmin$ increases (compare Fig.~\ref{Fig_BLS_butterfly}(b) with (a)). 
The relaxation frequency $\wrx$ of spin waves with $k \rightarrow 0$ as a function of $\jc$ reads\cite{Melkov_Book,Lvov_Book}
\begin{equation}
\wrx(\jc) = \dfrac{\pi}{2}\dfrac{\gamma^2}{f_\mathrm{sw}} \cdot \mu_0\Ms(\jc) \cdot \mu_0\hmin(\jc),
\label{eq:wR}
\end{equation}
and the relative variation of $\wrx$ is given by
\begin{equation}
\dfrac{\Delta \wrx(\jc)}{\wrx(0)} = \dfrac{\wrx(\jc)-\wrx(0)}{\wrx(0)}
= \dfrac{\hmin(\jc) \Ms(\jc)}{\hmin(0) \Ms(0)} -1.
\label{eq:wR-change}
\end{equation}

Figure~\ref{Fig_shift_omega}(b) shows the dependence of the relative variation of $\wrx$ as a function of $\jc$. These values are obtained according to Eq.~\ref{eq:wR-change} using the $\hmin(\jc)$ values from Fig.~\ref{Fig_shift_omega}(a), as well as Eq.~\ref{eq:def-b} and the fitting parameter $b$ for $\Ms(\jc)$. The relative variation of the relaxation is proportional to the current within the error bars, which is illustrated by the linear fit in Fig~\ref{Fig_shift_omega}(b). $\wrx$ is changed by approximately \units{7.5}{\%} for the highest applied current densities of \units[\jc=]{\pm 5 \cdot 10^{10}}{A/m^2} in our experiment. In order to estimate the required critical current density for the complete compensation of damping, and for the triggering of auto-oscillations in our experiments, the extrapolation of the linear fit to $\wrx(\jc)=0$ is performed, yielding a value of \units[j_\mathrm{c,crit}^\mathrm{exp}\approx]{6.7 \cdot 10^{11}}{A/m^2}. This value agrees well with the results obtained with patterned structures,\cite{HamadehPRL113,ColletArxiv} and with the theoretically estimated value of \units[j_\mathrm{c,crit}^\mathrm{theo}\approx]{8.3 \cdot 10^{11}}{A/m^2} calculated on the basis of analytical expressions provided in Ref. \onlinecite{HamadehPRL113}. For this calculation, the parameters of our system indicated above, the transparency of the YIG/Pt interface\cite{ZhangNPHYS11} \units[T=]{0.12}{} (estimated for a spin diffusion length of \units[\lambda=]{3.4}{nm}) and a spin Hall angle of \units[\theta_\mathrm{SH}=]{0.056}{} are used.\cite{Rojas-SanchezPRL112} Heating effects are neglected.
In addition, reference measurements are performed in a geometry with $\jc$ parallel to $\Happl$, so that the SHE is eliminated. As a result, no variation of $\wrx$ with $\jc$ is observed within the error bars (not shown).

In summary, the threshold of the parametric excitation of spin waves in a macroscopic YIG/Pt bilayer is detected by BLS spectroscopy. A change of the damping in the YIG due to SHE-STT is observed when a dc current is passed through the adjacent Pt film. Based on our calculations, including current induced Joule heating and Oersted fields, the damping is found to change linearly with the current. The linearity suggests that the role of nonlinear multi-magnon scattering processes is negligible for the damping variation in the analyzed range (below $10 \%$). The maximum used current density of \units[\jc=]{\pm 5 \cdot 10^{10}}{A/m^2} results in a damping variation of $\pm 7.5 \%$. Thus, the complete damping compensation in the system is estimated at \units[j_\mathrm{c,crit}^\mathrm{exp}\approx]{6.7 \cdot 10^{11}}{A/m^2} which is in agreement with the theoretical estimation.

\begin{acknowledgments}
This research has been supported by the EU-FET grant InSpin 612759 and by the Ukrainian Fund for Fundamental Research. D.A.B. has been supported by a fellowship of the Graduate School Material Sciences in Mainz.
\end{acknowledgments}

\end{document}